# Cellular Communications on License-Exempt Spectrum: A Tutorial

B. Ren, M. Wang, JJ Zhang, W. Yang, J. Zou, and M. Hua

## ABSTRACT

A traditional cellular system (e.g., LTE) operates only on the licensed spectrum. This tutorial explains the concept of cellular communications on both licensed and license-exempt spectrum under a unified architecture. The purpose to extend a cellular system into the bandwidth-rich license-exempt spectrum is to form a larger cellular network for all spectrum types. This would result in an ultimate mobile converged cellular network. This tutorial examines the benefits of this concept, the technical challenges, and provides a conceptual LTE-based design example that helps to show how a traditional cellular system like the LTE can adapt itself to a different spectrum type, conform to the regulatory requirements, and harmoniously co-exist with the incumbent systems such as Wi-Fi. In order to cope with the interference and regulation rules on license-exempt spectrum, a special medium access mechanism is introduced into the existing LTE transmission frame structure to exploit the full benefits of coordinated and managed cellular architecture.

## INTRODUCTION

The advent of smart mobile devices triggered the constant push for higher data rates. There is a strong trend of media consumption moving toward the mobile devices. Competitive market pressures continue to challenge wireless connectivity operators to create innovative technologies to deliver ever-growing volumes of mobile data, and to meet the ever-increasing demand for faster mobile data services in an already scarce radio spectrum. Today's wireless cellular network, like the LTE network, is already operating at a very high spectral efficiency, leaving little margin for further practical and cost-effective improvements. Small cells are believed to play a pivotal role in reaching more ambitious data rate for today's mobile applications through increasing the frequency reuse or the area spectral efficiency [1]. Currently, cellular small cell networks are unexceptionally operating on the licensed bands. Thus, the network capacity is ultimately upper-bounded by the availability of these licensed bands. Therefore, a key element to materialize the full potential of small cells is the introduction of a new technology that allows the small cells to operate beyond the limited licensed spectrum.

The Federal Communications Commission (FCC) uses several mechanisms to make spectrum available for wireless services through licensed and license-exempt spectrum. Licensed spectrum allows for *exclusive* use of particular frequencies or channels in particular geographic locations. Whereas in spectrum that is designated as license-exempt or unlicensed, users can operate without an FCC license but must comply with the constraints (e.g., the maximum transmit power limit) imposed by the FCC's regulations. Users of the license-exempt bands do not have exclusive use of the spectrum and are subject to interference. In certain regions, such as the European Union and Japan, the "listen-before-talk" (LBT) rule is enforced for better coexistence among different wireless systems (e.g., Wi-Fi, Bluetooth) that operate on the same unlicensed band. The LBT medium access rule requires that a transmitter wait for its turn if there is evidence that another transmitter is using the channel. A process called *clear channel assessment* (CCA) is used to determine if the channel is available for transmission.

Cellular systems and WLAN systems employ fundamentally distinctive architectures to cope with different channel properties, and to achieve different goals. Cellular systems operating on licensed spectrum are characterized by high spectral efficiency, reliable and predicable data service performance, and mobility, whereas most WLAN systems on license-exempt bands are typically cost-effective and flexible in deployment but are often spectrally inefficient and lack of quality of service (QoS) control. A natural choice could just be to extend LTE into license-exempt spectrum which aggregates both licensed and unlicensed bands to provide a seamless extension of a larger LTE network, allowing for seamless flow of data between licensed and license-exempt spectrum with the same technology through a single core network that employs the same authentication, operations and management systems, and the same acquisition, access, registration, paging and mobility procedures. This means reduced overhead, higher system performance, and strengthened overall network capacity. The question is then how to apply the cellular technologies to license-exempt spectrum to take advantage of the widely available, bandwidth-rich licensed-exempt bands; and how to offer *reliable* mobile services in *unreliable* license-exempt spectrum.

Indeed, extending today's cellular systems into the license-exempt band poses several serious technical challenges. In a cellular system, the network has the right of exclusive use of the spectrum. Therefore, the utilization of radio resources is guaranteed and there is no threat of *uncontrolled* interference. This allows the transmission and reception to be organized into a highly efficient frame structure that is <u>continuous</u> and follows a <u>deterministic</u> timing. With this *frame-based* structure, a network can continuously utilize the resources and efficiently manage them without the need for monitoring the channel activities or yielding to the traffic from other systems. The only interference is from its "friendly", cooperative neighboring cells belonging to the same network, which is mitigated through network planning [2] or coordination among cells [3], [4]. On the contrary, in a license-exempt band there is no guaranteed use of resources. Resources are actually being used in a competitive fashion via LBT, e.g., the Distributed Coordination Function (DCF). As a result, the transmission/reception structure is not fixed or deterministic but *opportunity-driven*, which is referred to as the *load-based* structure. It is hard to enforce an LTE-alike transmission structure in unlicensed spectrum and stringent QoS requirements that are essential to the cellular service are difficult to be ensured. The incompatibility of these two medium access mechanisms and transmission structures, and the different regulatory rules are therefore the key issues to the problem, and hence the focus of this tutorial.

Complementing the cellular system with unlicensed spectrum is increasingly considered by cellular operators as a complementary tool to augment their wireless services and solutions. In 2014, an LTE-U Forum was created by Verizon, in conjunction with Alcatel, Ericsson, Qualcomm, and Samsung. The motivation is to enable LTE to utilize the vast amount of available spectrum in the 5GHz band. As depicted in Figure 1, there is up to 500 MHz of spectrum at 5 GHz currently mainly used by Wi-Fi.

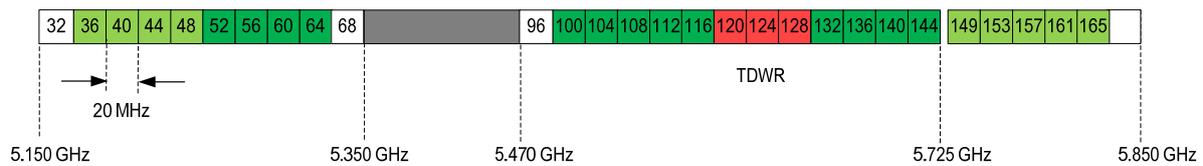

Figure 1 Channelization of the 5 GHz license-exempt band. The green-colored channels are unlicensed spectra with minimum bandwidth of 20 MHz that are currently mainly used by Wi-Fi, of which the channels in dark green requires the dynamic frequency selection spectrum-sharing mechanism to ensure coexistence with radar systems. TDWR channels (Channels 120, 124, and 128) are Terminal Doppler Weather Radar channels that are not allowed to be used for other purposes.

In its current stage, LTE-U Forum focuses on markets *without* LBT requirements for quick deployment without modifying LTE PHY/MAC standards. The direct deployment of LTE in the regions with the LBT requirements is prohibitive for many reasons. One of the simple reasons is that some LTE reference signals, such as CRS, have to be transmitted in a time-contiguous pattern even in the absence of data traffic, which violates the LBT channel occupancy time requirement that prohibits continuous transmission and impose limits on the maximum duration of a transmission burst as will be discussed in detail in this tutorial. Several mechanisms are proposed for fair and friendly coexistence of LTE with Wi-Fi systems operating in the 5 GHz unlicensed spectrum. For example, the channel selection approach simply looks for a cleanest channel that no Wi-Fi activity is present. It monitors the status of the channel on an on-going basis, and selects and switches to a more suitable channel if needed. This carrier selection scheme is used to avoid co-channel operation with Wi-Fi systems on a relatively slow time scale. In the event that no clean channel is available, this algorithm shares the channel with Wi-Fi systems based on 10s to 100s of msec carrier sensing of co-channel Wi-Fi activities.

Most recently, 3GPP has approved a work item to extend the LTE system into unlicensed spectrum in regions with or without LBT requirements [5]. Extension of the LTE to the license-exempt band (e.g., the 5GHz band) has formally started as part of Release 13 [6] (and the references therein). The study is focused on "license-assisted access" (LAA), in which the access to unlicensed spectrum via a secondary component carrier is assisted by a primary component carrier on licensed spectrum [7]-[8].

There is also a possibility to deploy LTE on unlicensed spectrum without the assistance from the licensed spectrum (i.e., the stand-alone model). This use case is challenged and not yet well received due to the concern that the performance gain of LTE over Wi-Fi on unlicensed spectrum without the assistance from licensed spectrum is limited, whereas the cost of an LTE modem is much higher than Wi-Fi.

In this tutorial, we utilize a conceptual design of a cellular system framework, henceforth referred to as LTE-C (**C**onceptual), to help better demonstrate the concept and the ideas behind [5], and the potential impact of the regulatory rules on the traditional LTE transmission structure. LTE-C is comprised of the traditional LTE-A (LTE Advanced) system, and a new LTE system that operates on the license-exempt spectrum, henceforth referred to as LTE-u (**u**nlicensed, a lower case of "u" is used in distinguishing it from the LTE-U Forum). We will also use this design to show how to implement LBT on an LTE frame-based transmission structure for more coherent interworking between licensed and unlicensed spectra, and between neighboring cells. Without loss of generality and for ease of discussion, we focus only on the downlink.

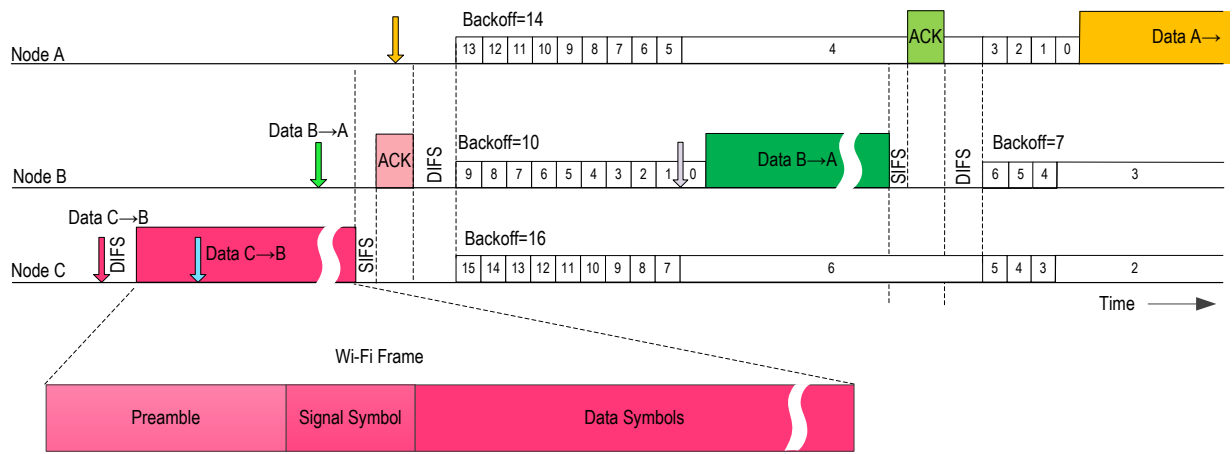

Figure 2 Illustration of Wi-Fi transmission pattern using DCF with a contention window length of 32. An arrow indicates the arrival of application data for transmission. It is evident that the transmission is dependent on the outcome of the DCF; there is no deterministic transmission timeline. For this simplified example of DCF, Wi-Fi Node C has data ready for Node B. Since there is no ongoing traffic in the channel, Node C transmits data immediately after the channel is assessed to be clear for a period of DIFS. Upon a successful reception of the data frame from Node C, Node B acknowledges with an ACK frame after the *short inter-frame space* (SIFS) time period. Since SIFS is defined to be less than DIFS, it preempts other nodes, allowing an immediate delivery of ACK to the transmitter. During this period, Nodes A and B, as well as Node C have new data ready for transmission. However, they all have to wait until the ongoing traffic finishes. Once the channel has been cleared for the DIFS time period, each node has to start a back off timer, uniformly distributed within a contention window of length 32, 14 for Node A, 10 for Node B, and 16 for Node C for example. The timer for Node B expires first and Node B transmits. Node A receives and acknowledges. The timers for Nodes A and C freeze until ACK finishes, and resume after the channel is cleared for the DIFS period of time.

## WI-FI AND LTE MEDIUM ACCESS AND TRANSMISSION STRUCTURES

In this section, we briefly review the load-based Wi-Fi and frame-based LTE transmission structures.

### Wi-Fi Medium Access and Transmission Structure

Most wireless systems deployed in license-exempt bands employ carrier sense multiple access (CSMA) as the basis for LBT medium access control (MAC). The most commonly used MAC is the DCF, employed by Wi-Fi systems, which is based on the CSMA with collision avoidance (CSMA/CA) scheme [9].

The principle of DCF can be best described using the illustration in Figure 2. If the channel is sensed idle for a specific duration, i.e., the *distributed inter-frame space* (DIFS) period of time, the node transmits. Otherwise the node continues monitoring until the channel is idle for DIFS. At this time, the node generates a *random* backoff timer, uniformly distributed within a contention window. The additional random sensing time helps avoid potential collisions, which may happen when two or more nodes are simultaneously waiting on the channel to be cleared. The backoff timer is decremented as long as the channel is idle but remains "frozen" when a transmission is detected, and reactivated after the DIFS period of time as soon as the channel is assessed to be free. A node refrains itself from transmission until the backoff timer expires.

We can now see that DCF tries to ensure that only one transmission is present (in range) in a channel at a time and each node has a fair share of the channel (via the random backoff). The channel use for each node at a particular time is not guaranteed however. Consequently, there is no deterministic timing structure for transmission, reflecting the random and contentious nature of communications in unlicensed spectrum. Reliable services and efficient resource usage are typically hard, if not impossible, to achieve.

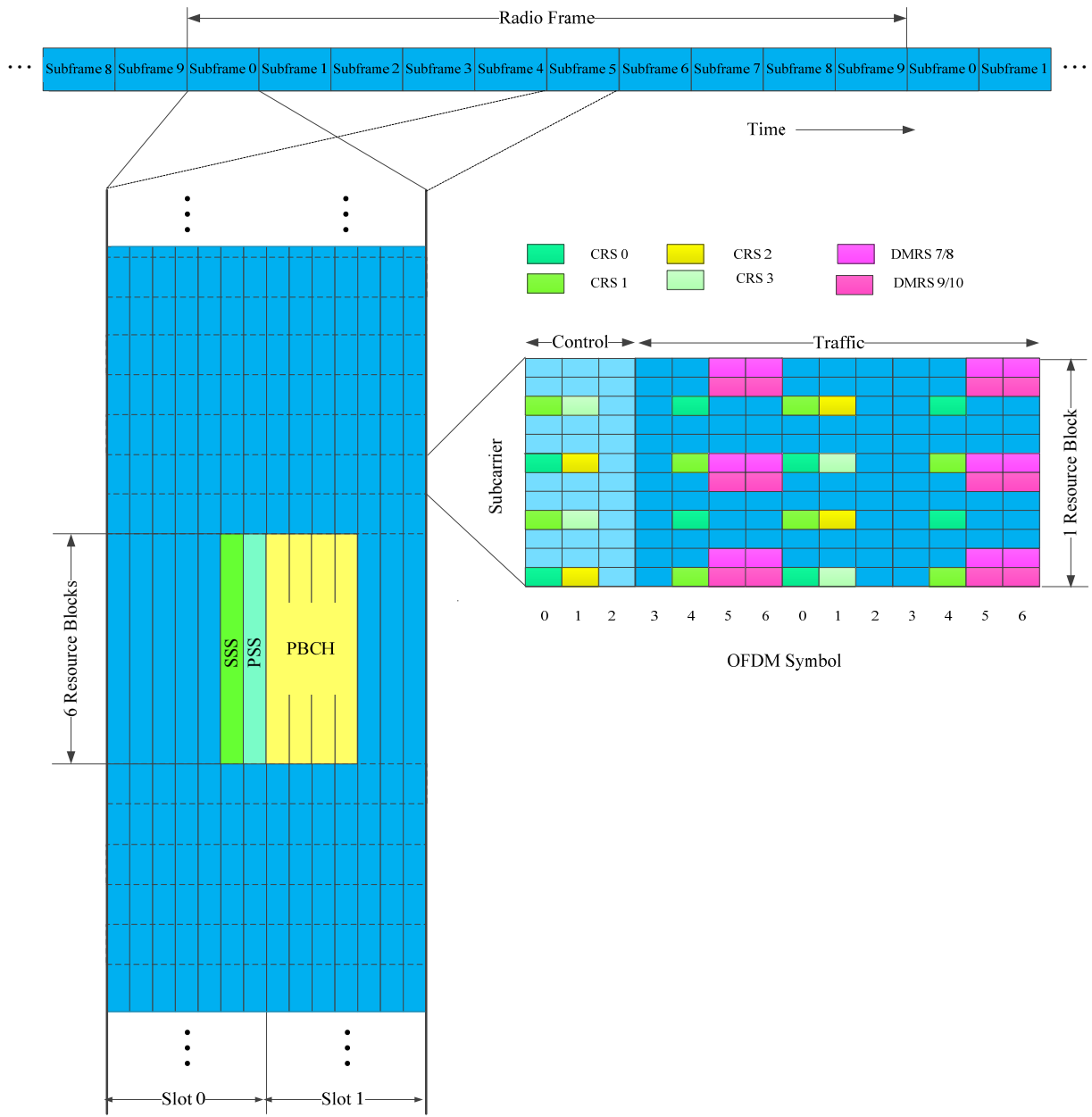

Figure 3 LTE downlink frame structure diagram.. In the frequency domain, resources are grouped in units of 12 subcarriers such that one unit of 12 subcarriers for a duration of one slot is termed a *resource block*. The basic transmission unit for data, i.e., the smallest resource unit that can be scheduled for transmission on the PDSCH is the resource block pair which consists of two resource blocks that last for duration of one subframe, thereby comprising 168 resource elements (REs). Signals like the Primary Synchronization Signal (PSS), the Secondary Synchronization Signal (SSS), and the Physical Broadcast Channel (PBCH) are also present on downlink in Subframe 0 (and 5 – PSS/SSS only), occupying the central six resource blocks. Combined, they provide transmission bandwidth, cell ID group, and timing for the radio frame, subframe, slot, and OFDM symbol for initial system acquisition and tracking. The system information contained in PBCH is repeated for four radio frames and updated every four radio frames (40 ms), enabling incremental redundancy decoding for receivers. CRS and DM-RS are also shown on a resource block pair. The first few OFDM symbols (maximum three) in Slot 0 are used for the downlink control channel (PDCCH).

This property is very different from that of communications in a licensed spectrum, consequently resulting in a very different transmission structure from that of a cellular system, as we will see in the next section.

A Wi-Fi frame consists of a preamble, a signal symbol and multiple data symbols as illustrated in Figure 2. The Wi-Fi preamble is a special waveform designed for Wi-Fi signal identification, AGC refinement, timing and frequency synchronization, and channel estimation. The Wi-Fi preamble is particularly suited for Wi-Fi activity detection in a channel since waveform detection is more sensitive (10-20 dB more sensitive) than the simple energy detection. Furthermore, since the Wi-Fi frame does not have fixed timing, the preamble is crucial for a receiver to synchronize to the frame. The signal symbol following the preamble contains the information that includes the modulation type, code rate, and the total number of octets for the following data symbols. The data symbols carry MAC packets. Depending on the payload of the MAC packets, the frame can be really short like the ACK frame or long like the frame for user traffic.

Each Wi-Fi MAC packet also contains a transmission duration field for informing the neighboring nodes of the medium occupancy time (even although the packet may be destined just for a particular node). This is an amount of time that all nodes must wait if they receive it. A local timer of a neighboring node is updated after the node reads the duration value from the ongoing transmission. This node defers the carrier sensing procedure until the local timer expires. The problem with this mechanism is that a node has to decode the whole MAC packet in order to read the duration field, and a long frame can be interrupted even before its completion. A protection mechanism commonly used in Wi-Fi, called "Clear-to-Send (CTS) to Self", is created to alleviate this problem. A CTS-to-self frame is a standard CTS frame except that it is addressed to the transmitting node itself. Like the ACK frame, the CTS-to-self frame is very short (since it does not contain any user data) and is typically heavily protected with the lowest code rate. The corresponding length of CTS-to-self is only 44 $\mu s$, and is hence less likely to be interrupted than a regular traffic frame. Although this MAC packet is addressed to the sender itself as the name implies, it is meant for the neighboring nodes and is honored by all the nodes that can read it. A node transmits a CTS-to-itself frame right before transmitting the traffic frame. The duration field of the CTS-to-self packet contains the time of the following traffic frame, thereby providing more effective protection of the subsequent frame.

## LTE Medium Access and Transmission Structure

In LTE (and any other cellular system deployed on the licensed spectrum), any transmission has to follow a continuous stream of a deterministic frame structure, termed a *radio frame*, designed for best transmission spectral efficiency, QoS control, and inter-cell coordination. As depicted in Figure 3, an LTE radio frame consists of ten 1-ms *subframes*, each of which is further divided into two 0.5-ms *slots*. Each slot is comprised of seven OFDM symbols (each of length ~66.7 $\mu s$) [10].

There are two important types of reference signals associated with this frame structure: the cell-specific reference signal, e.g., CRS, and the user-specific reference signal, e.g., DM-RS. The CRS, *constantly* present on downlink, enables the demodulation of the downlink control channels (PDCCH) and generation of channel state information feedback from users, and most importantly, provides a continuous reference for the time and frequency tracking loops of a mobile device. The DM-RS, present on both downlink and uplink, is used by receivers for demodulating user traffic or control data.

The MAC in the base station (or eNodeB in LTE terminology) includes a dynamic resource scheduler that allocates physical resources on physical downlink shared channel (PDSCH) for data traffic. The scheduler takes into account the traffic volume, the QoS requirement, and the radio channel conditions when sharing the physical resources among mobile devices. For downlink data transmissions, the eNodeB transmits the PDSCH grant, i.e., PDSCH resource assignments and their modulation and coding scheme (MCS), on PDCCH, and the data packet on the PDSCH accordingly. The mobile device monitors its PDCCH in the control region to discover its grant. Once its PDCCH is detected, the mobile device decodes the PDSCH on the allocated resources using the MCS provided.

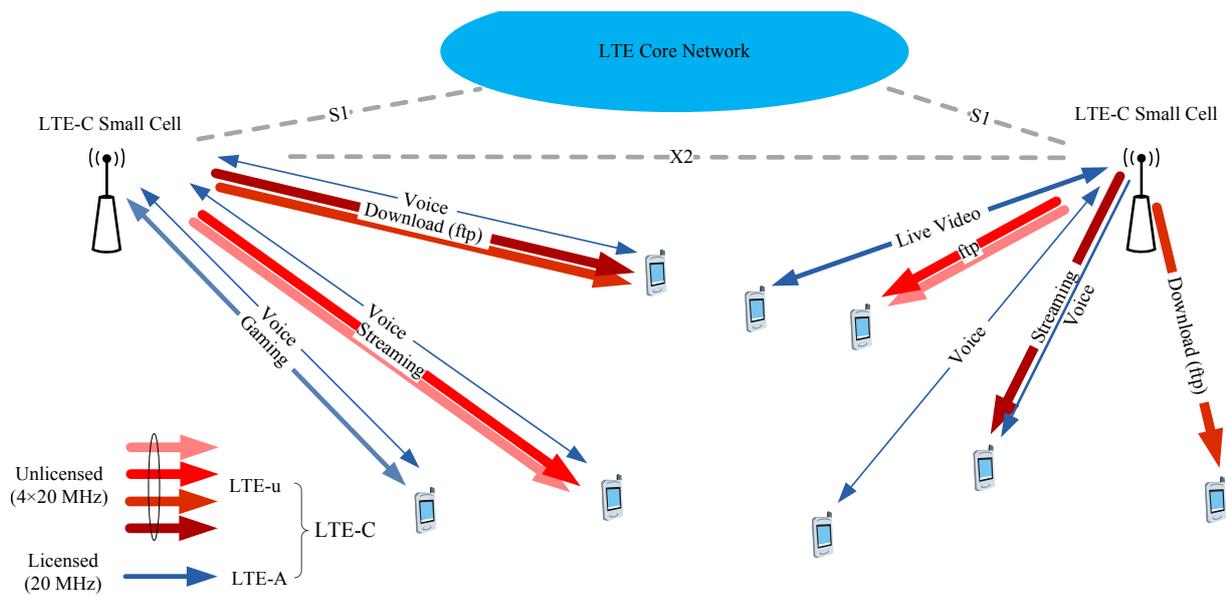

Figure 4 Illustration of a conceptual LTE cellular system (LTE-C) deployment scenario, similar to LAA [7], to exploit the full benefits of the centrally coordinated and managed cellular architecture, where data streams are aggregated and carried on both licensed band (e.g., 20 MHz using LTE-A air interface) and unlicensed bands (e.g., 80 MHz spectrum split into four 20 MHz bands using LTE-u air interface). QoS-crucial data services (e.g., voice, live video, and gaming) are delivered through LTE-A, whereas latency-insensitive data (e.g., ftp download) are delivered through LTE-u. The control signals/messages (not shown) are through LTE cellular infrastructure. In essence, LTE-C leverages the large number of small cells to work as a unified LTE network to efficiently exploit both licensed and unlicensed spectrum bands.

We observe that there is a clear distinction between LTE and Wi-Fi in terms of medium access and transmission pattern. For LTE, only one system (the system that owns the spectrum) is allowed to use the spectrum. As such, the MAC of the system can continuously utilize the resources and efficiently manage them freely without the need for monitoring the channel activities or yielding to the traffic from other systems. This results in a most resource-efficient transmission pattern that follows a continuous stream of a deterministic frame structure – there is no disruption for LBT. The continuous transmission structure also allows for the contiguous transmission of reference signals, such as CRS, providing continuous time and frequency synchronization references for receivers. Preambles are therefore not needed before each transmission. As for Wi-Fi, there is no guaranteed use of resources on unlicensed spectrum. Different Wi-Fi systems compete for resources following the LBT rule, e.g., DCF. As a result, the transmission/reception does not follow a fixed time frame and contiguous transmissions of reference signals for receiver time and frequency tracking is not possible. A preamble is thus necessary for every Wi-Fi transmission.

# LTE-u Medium Access and Transmission Structure

One key aspect that enables a cellular system to operate with high reliability and at a high spectral efficiency is that the "heart beat" of the cellular system, i.e., the control signal is protected with not only high reliability but also guaranteed timing that is carved into the transmission structure of a cellular system. Neither guaranteed timing nor reliability is possible for the Wi-Fi DCF structure on license-exempt spectrum. Therefore, in the following design, we rely on the traditional cellular system (LTE-A) to provide the reliable delivery of control signals or messages between the network and the mobile devices. That is, the licensed spectrum is used for network control as well as other QoS-critical data services.

Figure 4 illustrates an exemplary deployment model for LTE-C. Given the power limits placed on the transmitters using the license-exempt spectrum, LTE-C would be naturally used in small cells. Typical deployment scenarios include indoor, outdoor and hotspot coverage locations. In this model, LTE small cells are the foundation of LTE-C, i.e., they serve as the anchors. The primary component carrier of the LTE-C is on the licensed spectrum employing the traditional LTE-A, which provides a reliable means for time-critical control message exchanges between the network and the mobile devices for resource scheduling of licensed and unlicensed bands among LTE-C users, and the support for coverage and mobility (handover between cells of mobile devices), whereas the unlicensed bands using the LTE-u air interface serve as secondary component carriers mainly for traffic transportation leveraging license-exempt spectrum to opportunistically offload the "best-effort" class of data traffic from the network. This configuration allows for exploitation of the ultra-wideband license-exempt spectrum for *aggressive* high rate data services while relying on the traditional cellular infrastructure on licensed spectrum for reliable control and high-QoS data services, as well as for coverage and mobility.

As aforementioned, in order to operate on the license-exempt spectrum, a transmitter is usually required to follow an LBT rule for interference avoidance in order to coexist with other systems (e.g., the incumbent Wi-Fi systems) operating on the same unlicensed band. We could simply adopt the Wi-Fi DCF mechanism for LBT similar to the one as illustrated in Figure 2. However, the adoption of the load-based Wi-Fi DCF inevitably causes timing misalignment between LTE-u and LTE-A as well as between LTE-C cells. On the other hand, implementing LBT on the frame-based cellular transmission structure is not straight forward due to the incompatibility between LBT and LTE transmission structure. The goal of the LTE-u design is then to implement an efficient the LBT mechanism under a *unified* LTE frame-based structure.

Figure 5 (a) depicts the LTE-u frame structure. In this design paradigm, one subframe [the Subframe 0 in this example] of an LTE radio frame is designated for handling the unlicensed band LBT operation for coexistence with other systems. The remaining subframes (Subframes 1 to 9 in this example) are dedicated to data traffic by default, which span over a 9 ms period. This way, the channel occupancy time falls within the range from 1 ms to 10 ms required by the regulation.

To maintain maximum efficiency in both spectrum and power usage as in a traditional cellular system, synchronization within the LTE-C network is retained in the design. That is, an LTE-u cell *strictly* follows the same transmission timeline within the network as in LTE-A. This is realized using the concept of *dynamic* LBT upon a *fixed* transmission framework to deal with the randomness in unlicensed bands while still

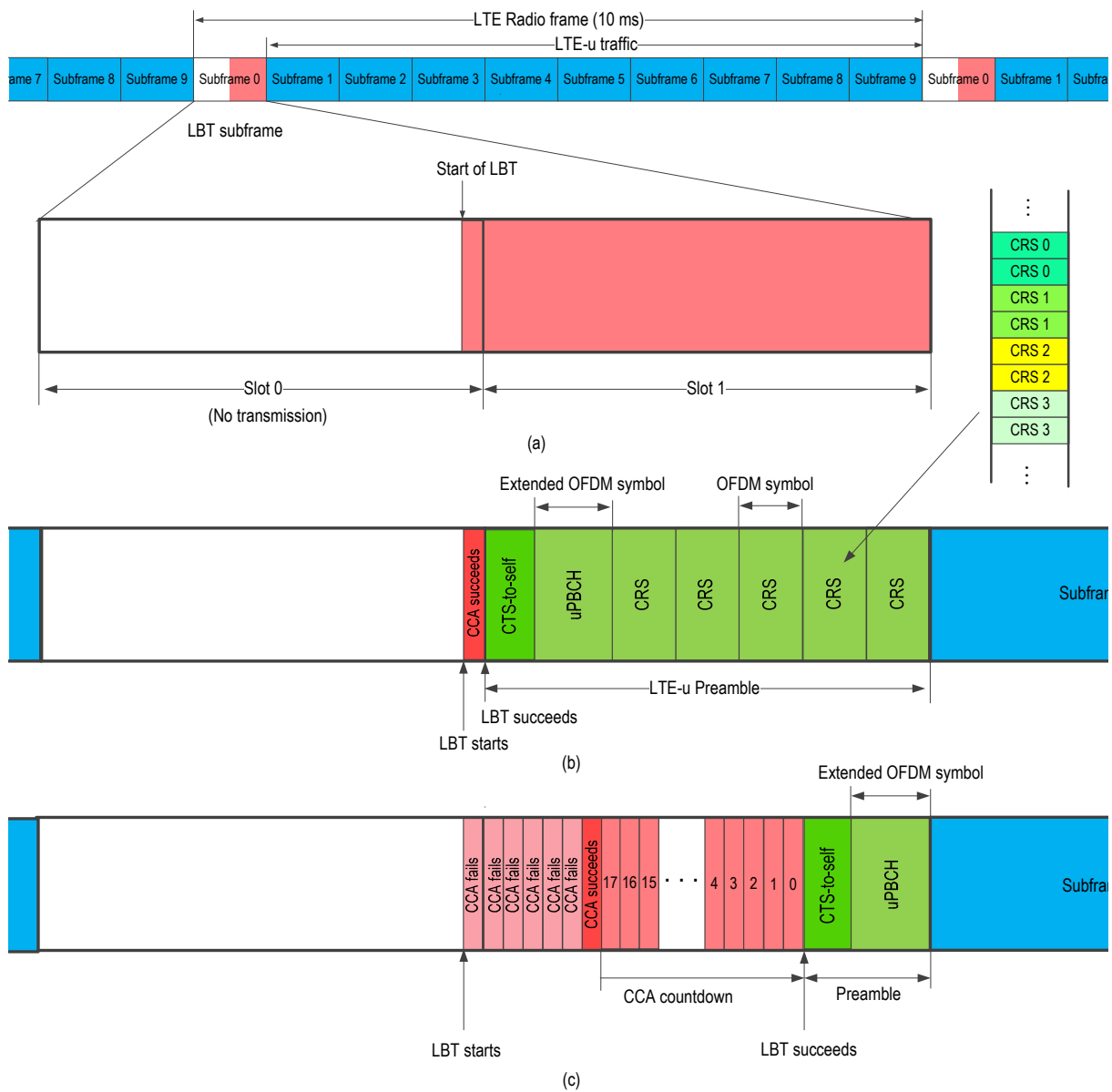

Figure 5 (a) The exemplary LTE-u frame structure, where Subframe 0 is devoted to LBT to cope with the interference and regulations in unlicensed spectrum, and Subframes 1 through 9 are reserved for traffic. (b) CCA succeeds before Slot 1 of Subframe 0; the preamble is transmitted *immediately* to secure the channel before the Subframe 1 starts. (c) Initial CCA fails and CCA continues until it succeeds; CCA countdown starts immediately after the first successful CCA. The initial value of the counter is randomly selected from 0 to 31 and decrements at every successful CCA. The preamble is led by a Wi-Fi CTS-to-self signal with a duration field indicating the transmission length including the duration of the following uPBCH [and CRS signal(s) if present] plus the active subframes (i.e., Subframes 1-9 in this example). It is most likely that the end of the CTS-to-self signal is not aligned with the OFDM symbol boundary; the length of the CP of the following uPBCH is extended to absorb the timing discrepancy. If the space between the end of CCA countdown is less than the length of CTS-to-self (44 μs) plus an OFDM symbol length, the preamble will extend all the way to the end of the first slot of the subsequent subframe.

maintaining a strict LTE frame-based timeline *synchronous with* LTE-A on licensed spectrum. This key concept will be introduced in the following sections.

As noted before, one subframe of an LTE radio frame is devoted to implementing the LBT mechanism in the LTE-u radio frame structure. The first half, i.e., Slot 0, of this LBT subframe is strictly reserved as the

"idle" period in which no transmission is permitted. The time duration of this idle period is hence 0.5 ms. The need for this idle period and the choice of the length is related to the minimum channel idle period requirement of the regulation and will soon become clear.

Like the Wi-Fi DCF described in the previous section, the proposed LBT mechanism is based on the CSMA/CA to ensure the smooth coexistence with other wireless systems such as the incumbent Wi-Fi. Referring to Figure 5 (b), LBT starts one CCA before Slot 1 of the LBT subframe, where the base unit of CCA is 25 $\mu s$, which is longer than the minimum CCA observation time of 20 $\mu s$ as required by the regulation. If the CCA is successful (i.e., no channel activity is detected), the preamble is transmitted immediately to secure the right to use the channel for the following nine subframes of the current radio frame. If the first CCA fails, CCA continues until it succeeds. At this time, a CCA counter is set off, whose initial value is randomly selected from 0 to, e.g., 31. The counter decrements in response to every successful CCA as illustrated in Figure 5 (c). The LBT ends when the counter expires. The preamble is then continuously transmitted in the remainder of Subframe 0 till the start of Subframe 1. Note that the finish of LBT does not guarantee the use right of the upcoming channel. In fact, the channel is up for grabs until a signal is injected into the channel. Since no transmission is allowed in Slot 0 of the LBT subframe, the earliest time that a preamble can be transmitted is the start of Slot 1. This is the reason that LBT starts one CCA before Slot 1.

In this design, the LTE-u preamble is led by the Wi-Fi CTS-to-self signal followed by uPBCH (LTE-u PBCH) and the optional CRS symbol(s). The duration field of the CTS-to-self signal includes the duration of the following uPBCH symbol [and CRS symbol(s) if present] *plus* the active subframes (i.e., Subframes 1-9 in the example of Figure 5). Since the CTS-to-self is a Wi-Fi signal as noted earlier, it can be read by any Wi-Fi node. As such, in the eyes of a Wi-Fi system, an LTE-u system is no different than a regular Wi-Fi system, and hence the medium access time as indicated in the CTS-to-self signal will be honored by a Wi-Fi system, thereby allowing for better protection against Wi-Fi transmissions. Since an LTE-u system also honors the Wi-Fi CTS signal during LBT, the protection naturally works both ways, ensuring smoother coexistence between LTE-u and Wi-Fi systems.

The uPBCH contains the identifcation (0 – 503) of the current cell that can be read by all the LTE-u systems. It is most likely that the end of the CTS-to-self signal is not aligned with the OFDM symbol boundary; the CP length of the uPBCH is extended for transition to the OFDM symbol timing of the LTE frame structure. The CRS symbols are optional, and are used to fill the gap (if any) between the end of the uPBCH symbol and the start of the next subframe. In the example of Figure 5, four CRS ports are provided for wideband channel state estimation. If the space between the end of CCA countdown and the start of the next subframe is less than the minimum length of the LTE-u preamble [CTS-to-self (44 $\mu s$) plus an OFDM symbol length (uPBCH)], the LTE-u preamble will extend all the way to the end of the slot of the following subframe to time-align with the slot boundary. When that happens, only 8.5 subframes (17 slots) can be used for user traffic.

In general, LBT may not end before the start of Subframe 1, depending on the interference (e.g., Wi-Fi traffic) pattern. In this case, the LBT extends into the following subframe(s) or even the following radio frame(s). As the example shown in Figure 6, the LTE-u system is preempted by a Wi-Fi system during the CCA countdown, causing the extension of the LBT well into the radio frame, i.e., Subframe 4 of the radio frame. As a result, the next LBT subframe is Subframe 4 of the next radio frame. Therefore, the LBT frame is not necessarily Subframe 0.

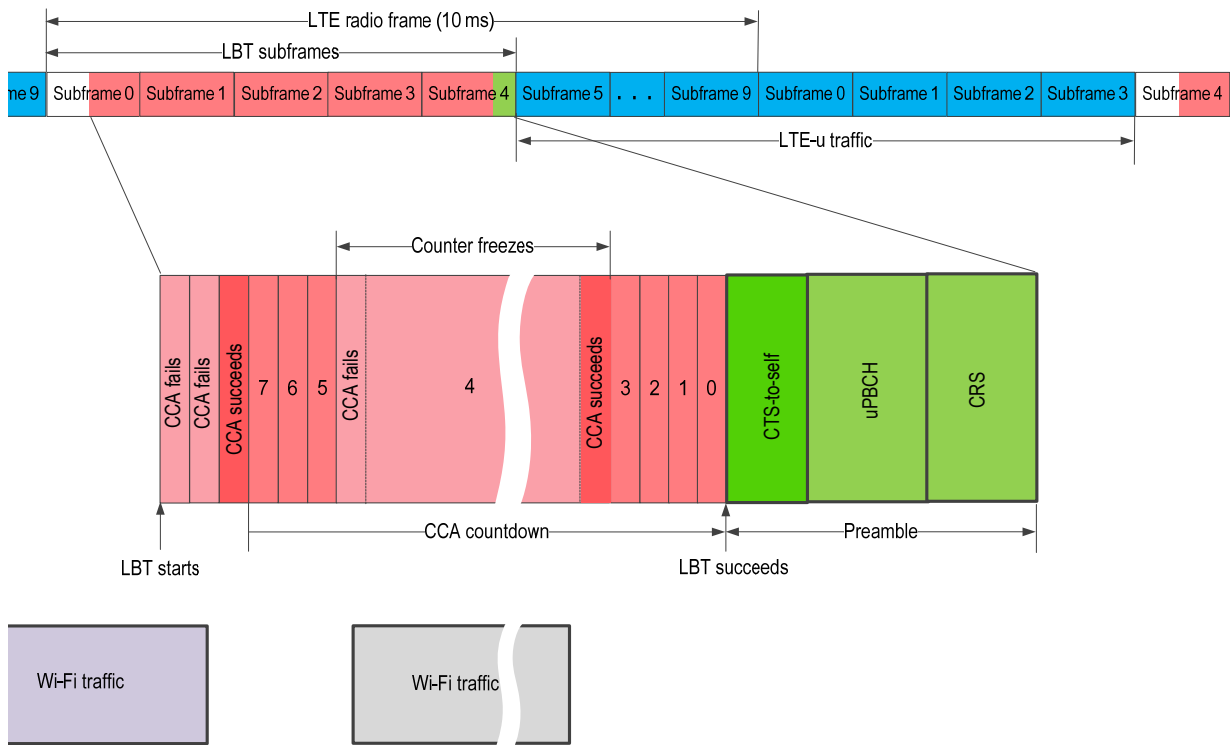

Figure 6 An example of the LBT of the LTE-u system that spans across multiple subframes. Initial CCA of the LBT started in subframe 0 fails as a result of the ongoing traffic from a Wi-Fi system. The LTE-u system is preempted by another Wi-Fi system during the countdown, as a result of a failed CCA (on count 4). The counter freezes until a successful CCA. As soon as the LBT succeeds, the preamble is transmitted to secure the use of the channel for the next 9 subframes. The next LBT starts from subframe 4 in this example. The preamble is led by a Wi-Fi CTS-to-self signal with a duration field indicating the channel time from the end of the CTS signal to the end of the following 9 subframes.

The initial value of the counter, ranging from 0 to, e.g., 31, is selected by a predetermined pseudo-random sequence with a given seed. At different radio frames, an LTE-u system obtains different countdown values from the pseudo-random sequence, ensuring fairness among different systems. Furthermore, with the anchor in the licensed channels, the interference between the cells and handover of mobile devices between cells are continuously and seamlessly managed. This design eliminates the core limitation to the mobility in Wi-Fi systems.

Note that, a successful LBT only warrants the right to use the current radio frame, i.e., the following nine subframe worth of channel time. A cell must give up the use of the channel as soon as the nine subframes elapse, and re-compete for the use of another radio frame worth of medium time. It is apparent that, under this design, the maximum channel occupancy time is 9 ms + 0.5 ms = 9.5 ms, which is still within the 10 ms range. The 0.5 ms idle period is hence more than 5% of the maximum channel occupancy time as required by the regulation.

## CONCLUSION

Compared to the Wi-Fi system, the traditional cellular system is highly-spectrally efficient but at the same time extremely vulnerable to uncontrolled interference. Direct deployments of a cellular system in a license-exempt spectrum are thus prohibitive. The key component that creates these differences is their unique transmission structures crafted to suite the unique channel properties of different types of spectrum. This tutorial explains these fundamental differences due to the very different interference properties and regulation rules existing in these types of spectrum. This tutorial describes a conceptual LTE system (LTE-C) that enables the operation of LTE in license-exempt spectrum. An example system is utilized for illustrating how a traditional cellular system can be "mutated" to operate on a different type of spectrum complying with the regulations, and most importantly, co-existing with other systems on the same band. Interference is the primary issue on unlicensed spectrum, especially between two different systems. There are three key components employed in LTE-u that provide the capability of interference/jamming avoidance, and fair sharing of the spectrum with other wireless systems, particularly the Wi-Fi system, in a cognitive fashion: (1) The use of a special CSMA/CA-based LBT mechanism. Since CSMA/CA is also the foundation of Wi-Fi DCF, the coexistence of LTE-u and Wi-Fi is hence not only natural but also efficient. (2) Although CSMA/CA is random and asynchronous in nature, the design lends itself well to the more efficient deterministic synchronous frame-based transmission structure within the LTE network, allowing coherent interworking between LTE-u and the anchor carrier LTE-A of LTE-C as well as neighboring LTE-C cells. (3) The inclusion of the Wi-Fi CTS-to-self signal as the preamble of LTE-u. Since CTS-to-self is a Wi-Fi message commonly used by Wi-Fi systems to control interference, it serves seamlessly as a "common language" for interference coordination between LTE-u and Wi-Fi systems, which further improves the co-existence between these two very different systems.

## ACKNOWLEDGEMENT

The authors would like to thank the Editor and the reviewers for their excellent comments.